\documentclass[a4paper,12pt]{article}

\usepackage{array}
\usepackage{multicol}
\usepackage{wrapfig}
\usepackage{graphics}
\usepackage{epsfig}

\begin{document}
\Huge
\title{Spin--dependent recombination -- an electronic readout mechanism 
for solid state quantum computers}
\author{Christoph Boehme\footnote{electronic mail: boehme@hmi.de}, Klaus Lips \\
Hahn--Meitner--Institut Berlin, Kekul\'estr. 5, D-12489 Berlin, Germany}
\date{\today}
\maketitle
\begin{abstract}
It is shown that coherent spin motion of electron--hole pairs localized in band 
gap states of silicon can influence charge carrier recombination.
Based on this effect, a readout concept for silicon 
based solid--state spin--quantum computers as proposed by Kane is suggested. The 
$^{31}$P quantum bit (qbit) is connected via hyperfine coupling to the spin of the 
localized donor electron. When a second localized and singly occupied electronic 
state with an energy level deep within the band gap or close to the
valence edge is in proximity, a gate controlled exchange between the $^{31}$P 
nucleus and the two electronic states can be activated that 
leaves the donor--deep level pair either unchanged in a $|T-\rangle$-state
or shifts it into a singlet state $|S\rangle$. Since the donor deep level 
transition is spin--dependent, the deep level becomes charged or not, 
depending on the nuclear spin orientation of the donor nucleus. Thus, 
the state of the qbit can be read with a sequence of 
light pulses and photo conductivity measurements.
\end{abstract}


\clearpage
\normalsize
\section{Introduction}
The state of the art of classical computer concepts 
is rapidly approaching the physical limits as fabrication technology of 
metal oxide semiconductor logic is minimized to scales 
where quantum effects determine device properties. While these
natural limitations of classical electronics are the dead end for the 
development of conventional electronics, they open up possibilities for new alternative
concepts such as spintronics and quantum computing (QC)~\cite{awschal}. 
A concept for a silicon based solid state spin--quantum computer as outlined by 
Kane~\cite{Kane,Kane2}, combines the advantages of conventional 
semiconductor technology with regard to the high degree to which this technology 
has been developed and the fundamental concepts of QC, the massive parallel
processing of information by coherent quantum states. Kane's 
concept takes advantage of the two nuclear--spin energy eigenstates of $^{31}$P-donor 
nuclei which can be used as well--isolated (long relaxation times) quantum bits (qbits)
if they are embedded in a nuclear spin--free crystalline $^{28}$Si matrix. 
Interaction between these nuclear spin qbits can be controlled by electric fields from 
charged metal gates above and between the donor atoms which can selectively increase the 
hyperfine interaction between the localized electron donor states as well as 
the exchange interaction between electron donor states of different 
$^{31}$P-atoms~\cite{Kane,Kane2}. Before an implementation of the silicon based 
spin--QC is possible, many technological challenges have to be overcome among which
the problem of a single spin readout is particularly difficult.
In the original proposal~\cite{Kane}, the readout of nuclear spin states is done 
by charge measurements of the qbits' electronic shell which can
contain one or two donor electrons from adjacent $^{31}$P-atoms.
Recently, other proposals for the 
measurement of a single nuclear spin state have been made utilizing single electron 
transistors~\cite{Kane3} or spin--transport in combination with 
spin refrigeration/spin--readout devices~\cite{Kane2}. 

In this study, a concept is presented, which utilizes spin--dependent 
charge carrier recombination in silicon for nuclear spin measurements. 
Electronic transitions between localized, singly occupied 
\begin{figure}[t]
\begin{center}
\includegraphics{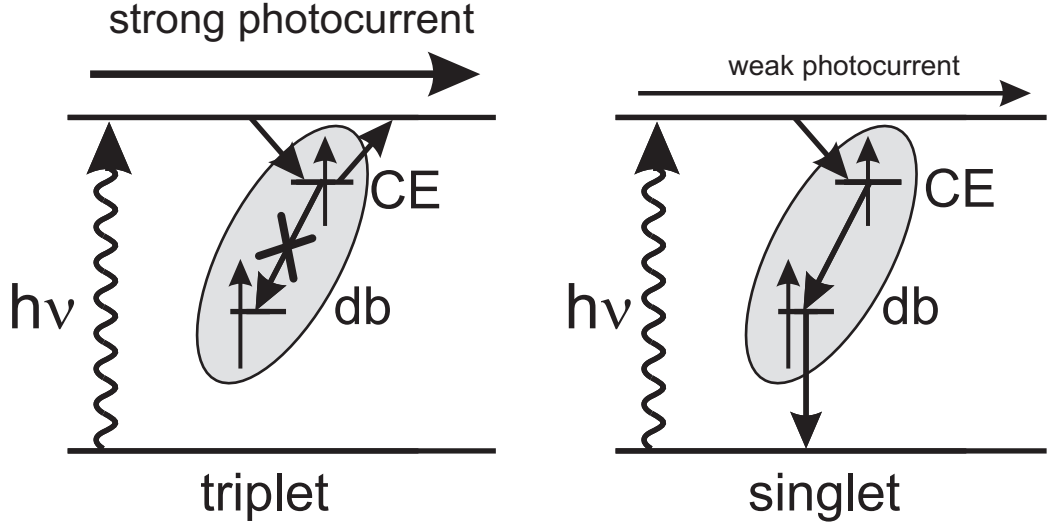}
\end{center}
\caption{\footnotesize{
Spin--dependent recombination in the 
picture of Kaplan et al. for the example of CE-db recombination in $\mu$c-Si:H: 
Since spin--orbit coupling is absent in silicon, the electron in the CE 
states can only charge the dangling bond, when the CE-db-pair state has singlet content.}
}
\label{fi1}
\end{figure}
paramagnetic band gap states in silicon have known to be spin--dependent 
since Lepine~\cite{Lep} discovered that electron spin resonant (ESR) changes of such 
defect states can change recombination rates and therefore photoconductivity. The
spin selection rules on these recombination transitions arise from the weakness of 
spin--orbit coupling in silicon which imposes spin--conservation. Thus, before a
transition into a singly occupied state can occur, the two electrons participating 
have to be in a spin state with singlet content. A model of spin--dependent 
recombination has been developed by Kaplan et al.~\cite{KSM}, who suggested, that 
spin--dependent recombination processes are always preceeded by a selective pair 
formation of the recombining electrons and holes. Figure~\ref{fi1} illustrates 
an example of such a process, which was observed by Kanschat et al.~\cite{Kan1,Kan} in
hydrogenated microcrystalline silicon ($\mu$c-Si:H). Due to the disorder of this material, 
shallow trap states exist close to the conduction band which can be singly occupied with 
excess conduction electrons (CE) at low temperatures. Beside these CE centers,
additional dangling bond defect states (db) exist which are excellent recombination 
centers. A formation of a spin--pair as described by Kaplan et al.~\cite{KSM} takes 
place, when a CE state in proximity of a db is occupied.
Similar to this example, spin--dependent recombination 
mechanisms exist in other silicon based materials such as dangling--bond recombination 
in amorphous silicon~\cite{stuke,stutz}, various donor--acceptor and 
donor radiation--defect recombination pathes in crystalline silicon (c-Si)~\cite{spae1,spae2}, 
in silicon devices~\cite{lips1,rong} and even at c-Si/silicon dioxide 
interfaces~\cite{stathis} where spin--dependent recombination occurs between stress 
induced traps and deep interface states. 

With regard to silicon QC, the selectiveness of these spin--dependent 
recombination mechanisms raises the question, whether they could be utilized
for a readout of electronic and hence also nuclear spin states, especially since 
$^{31}$P donors are known to show nuclear spin resonance influence on 
recombination~\cite{spae3}. Precondition for the information readout of coherent 
spin states with recombination processes is the ability of a given transition 
to reflect the coherence of the spin states involved, which means 
the spin--pair states that determine whether a recombination transition takes place or 
not must not fall into one of the four eigenstates, before a second transition, the 
actual electronic transfer takes place. This however has never been proven in the past
since the experiment used for the detection of all the processes mentioned above 
(often refered to as electrically detected magnetic resonance, EDMR) 
is a pure incoherent steady--state magnetic field--sweep experiment, where a constant 
current of light or electrically injected excess charge carriers is measured while a 
constant microwave radiation is imposed on the sample. When the magnetic field reaches 
the resonance of one of the involved centers, a change of the steady 
state recombination and thus a current change takes place. While EDMR shows that certain
magnetic centers have an influence on recombination, it fails completely to reveal any 
information about coherence times of the systems involved which are not only the 
spin--relaxation times of the respective spin centers but also electronic transition 
times such as recombination or dissociation of the spin pair. 

In the following an experiment is presented that shows how coherent spin precession of 
electrons in localized band gap states can govern recombination rates. Based on this 
observation we propose an architecture for a recombination based readout of 
$^{31}$P-qbits.

\section{Coherent spin motion effects on recombination}

Time--domain measurements of spin--dependent recombination (TSR) were carried out
on the CE-db mechanism in $\mu$c-Si:H, that is mentioned above. 
TSR is the electrical detection of pulsed ESR and therefore 
the time resolved equivalent of
\begin{wrapfigure}[28]{r}{85mm}
\begin{center}
\includegraphics{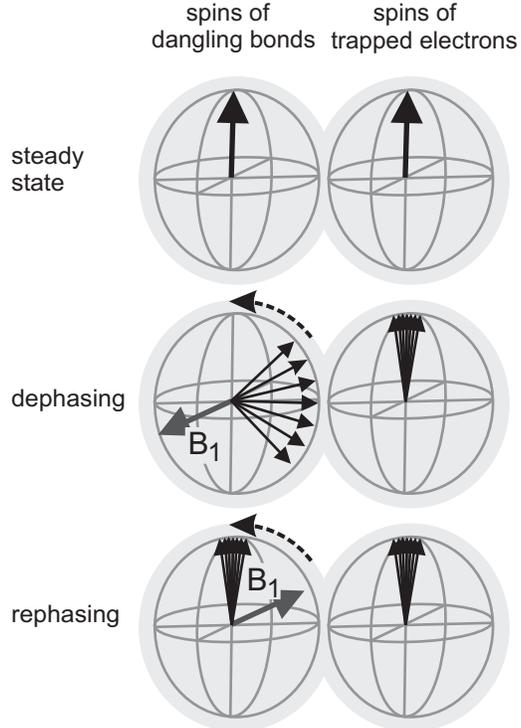}
\end{center}
\caption{\footnotesize{
The propagation of the CE-db pair ensemble illustrated with Bloch spheres in the
rotating frame picture. The three scetches correspond to the steady state, 
the moment of phase reversal at a time $t=\tau_{180^o}$ after the pulse 
began and the moment of phase recovery $t=2\tau_{180^o}$ for strongly distributed
Rabi--frequencies.}
}
\label{fi2}
\end{wrapfigure}
EDMR~\cite{Boe4}. 
It combines the advantages of EDMR with regard to the selective detection of distinct 
recombination processes and the advantages of pulsed ESR with regard to its 
ability to detect coherent spin-motion effects. 

The idea of the experiment carried out is to manipulate the steady state of 
the CE-db pair--ensemble with pulsed ESR on a nanosecond time scale and to observe the 
transient photocurrent response that is determined by the recombination rate through 
these centers. Excess charge carriers are generated by a continuous light source.
Since singlet states have only short lifetimes, a high density of
triplet states is present in the steady state when resonant microwave 
radiation is absent~\cite{Boe1}.

The orientation of the magnetic moments of $|T_-\rangle$--pairs is illustrated in 
fig.~\ref{fi2} in the rotating frame Bloch--sphere representation~\cite{SpiEch} of the two 
respective spins contained in the charge carrier pairs.
If a microwave that is in resonance with the dangling bond and 
polarized along the $x$--axis is switched on, Rabi--oscillations of the db spins take place. 
The influence of the microwave pulses on the CE centers is small due to their
different resonance frequencies. The Rabi--oscillation of the db spin would lead to an 
oscillating recombination rate which could be detected as an oscillation of the 
photocurrent, if the inhomogeneities present in $\mu$c-Si:H~\cite{Kan1} 
did not cause a rapid dephasing of the db ensemble within less than one oscillation 
period (illustrated in fig.~\ref{fi2}). 
This dephasing is still a coherent process, however, its effect on 
the recombination rate (strong attenuation of the Rabi induced oscillation) makes it 
indistinguishable from incoherence. Thus, a rephasing of the ensemble has to be induced, 
so that incoherence and dephasing becomes distinguishable. As illustrated in fig.~\ref{fi2},
a reversal of the direction of spin precession caused by a reversed $B_1$-microwave field
causes such a rephasing. The $B_1$--field reversal, done by a microwave phase change 
that is introduced at an arbitrary time $\tau_{180}$, causes a maximum recovery of the 
$|T_-\rangle$--density at time $2\tau_{180}$ and hence, a minimum of the 
recombination takes place. 
This minimum, which we call a recombination echo~\cite{Boe1} will only be detectable,
if incoherent processes are not faster than the time scale of the pulse sequence.
The detection is a proof of the coherence of the charge carrier pair's spin motion. 

Experimental details with regard to the $\mu$c-Si:H sample and the TSR setup
are identical to those of ref.~\cite{Boe4} execept that the laser intensity was 300mW, the 
sample temperature $T$ = 30K and the intensity of the coherent microwave radiation 500W. 

\begin{figure}[t]
\begin{center}
\includegraphics{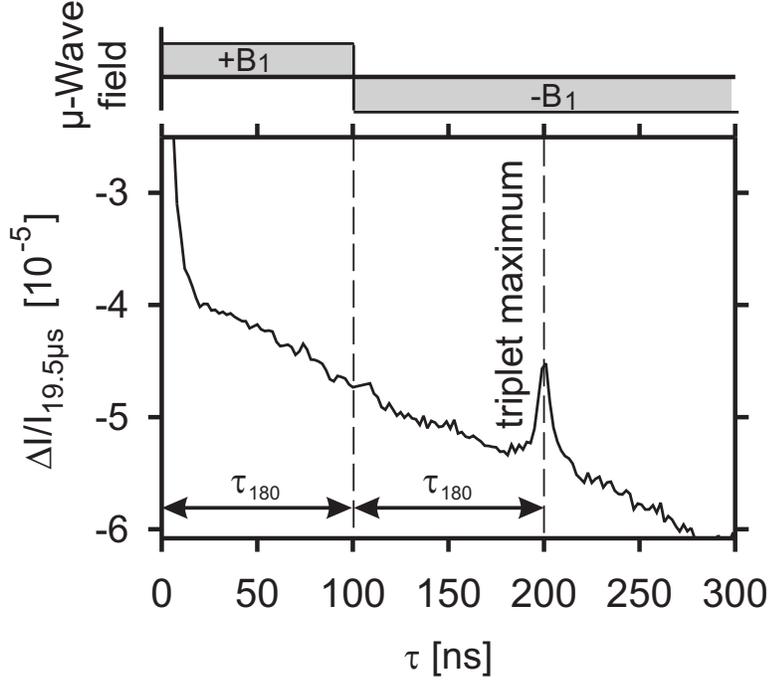}
\end{center}
\caption{\footnotesize{
The amplitude of measured current transients after resonant pulse sequences. After 100ns,
the polarization of the microwave field $B_1$ is reversed into its opposite direction, which 
causes a rephasing after a second period of 100ns. This rephasing after two periods of length 
$\tau_{180}$=100ns is indicated by the increase of the photoconductivity.}}
\label{fi4}
\end{figure}
\begin{figure}[t]
\begin{center}
\includegraphics{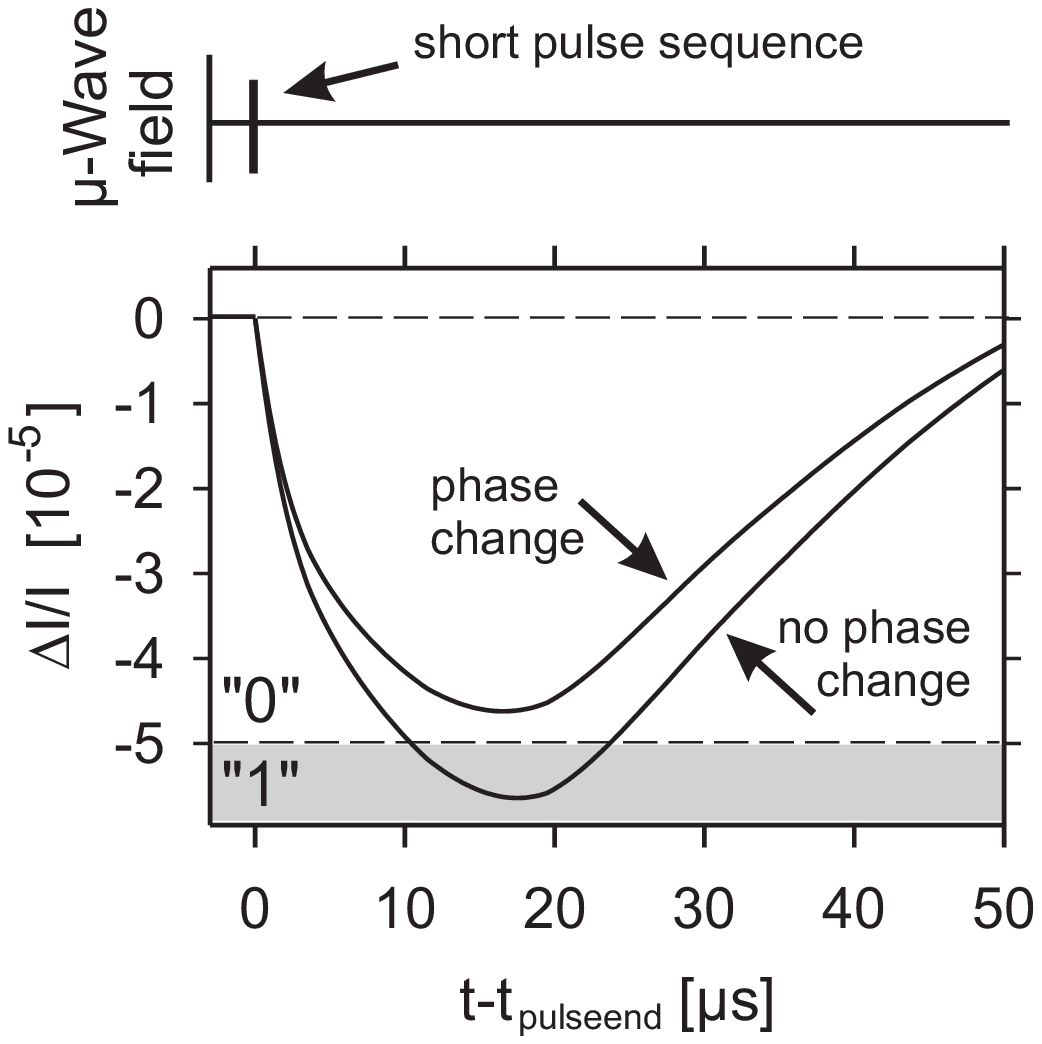}
\end{center}
\caption{\footnotesize{
Real--time photocurrent transients after resonant microwave excitation of equal length (200ns),
equal frequency (db--resonance) and equal intensity (P=500 Watt). The transient where a phase 
change was introduced after 100ns exhibits a smaller photocurrent quenching due to the higher 
triplet density among the CE-db pairs. When triplet states represent a binary information 
``0'' and singlet states represent ``1'', a current discriminator connected to the sample
can read out a spin state electronically.}}
\label{fi3}
\end{figure}

The experimental results are displayed in figs.~\ref{fi3} and \ref{fi4}. Note that 
the detection of fast photocurrent changes in the pA-range with constant offsets in 
$\mu$A-range is extremely difficult due to noise limitations. The inital decrease 
of the real--time photocurrent displayed in fig.\ref{fi3} is therefore due to the rise time 
of the detection setup. As explained elsewhere~\cite{Boe3,Boe4}, the amplitude of 
the exponential photocurrent relaxation back to the steady state is determined by the 
coherent spin state prepared during the ns-pulse sequence. Hence, a measurement of the 
real time signal at a certain time after the rise time versus the pulse length reveals the 
spin motion during the pulse sequence on a ns-scale. This pulse length dependence, measured 
at the photocurrent minimum ($t=19.5\mu$s) after the pulses, is displayed in fig.\ref{fi4}.
It clearly displays the photocurrent increase after a sequence of two 100ns long 
microwave pulses with equal $B_1$-field strength but opposite phase orientations.
The width of the echo displayed in fig.~\ref{fi4} turns out 
to be antiproportional to the $B_1$-field strength which proves the involvement 
of Rabi--oscillation in the recombination echo effect (not shown here).
The two real time transients displayed in fig.\ref{fi3} are recorded after 
resonant excitation of the db center.
In both cases, the total length and the power of the pulses were identical and only a 
phase change of 180$^o$ was introduced in one case. The smaller amplitude of one transient
is solely due to this phase change after 100ns which led to the rephasing of 
triplet states. An identification of singlet and triplet states with the two binary 
digits (labeled ``1'' and ``0'' for instance) used for information processing allows
a direct readout with current measurements.

\section{Spin--dependent recombination and nuclear spin measurements}

\begin{figure}[t]
\begin{center}
\includegraphics{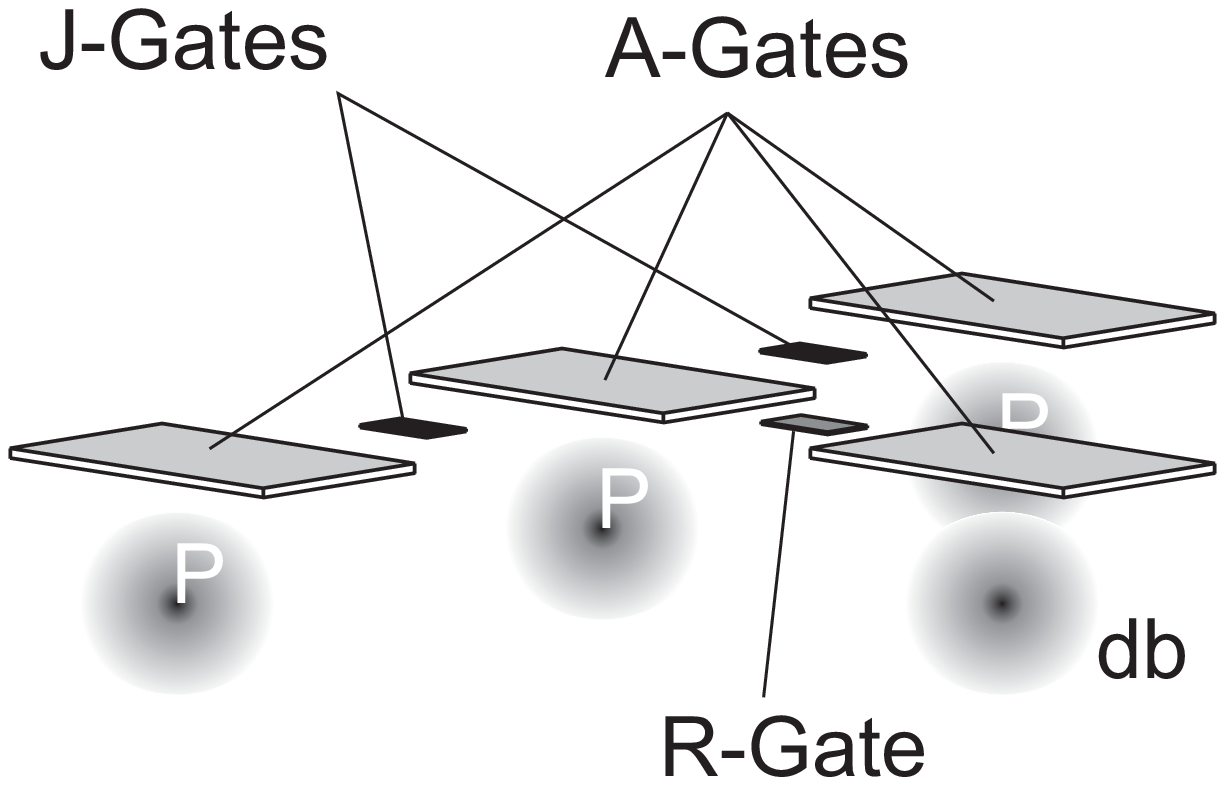}
\end{center}
\caption{\footnotesize{
Illustration of the readout setup. A point defect induces a paramagnetic deep level state
(here labeled db) in proximity to the $^{31}$P donor. The charge on a readout--gate (R-gate) 
controls the P-db-exchange interaction. When the P-db-pair is in singlet state after the 
R-gate is opened and the exchange is present, the P-electron can undergo a transition into 
the db state which charges the P center positively and the db-center negatively.}}
\label{fi5}
\end{figure}

Based on the observation above, a setup for a 
recombination readout of Kanes' silicon based quantum computer is proposed as 
illustrated in fig.~\ref{fi5}. The device proposed consists of a point defect 
that introduces a deep level paramagnetic state in proximity to the $^{31}$P-donor which 
is to be read out, a readout--gate (R-gate), and an additional A-gate above the defect. 
The latter is only necessary if the defect is induced by a deep level donor with 
nuclear spin ($I\neq0$). When the nature and the implementation of the paramagnetic deep level 
state will be discussed in the following section, it will be refered to as db in this 
section in consistency with the dangling bond used for the experimental demonstration 
discussed above. 

\begin{figure}[t]
\begin{center}
\includegraphics{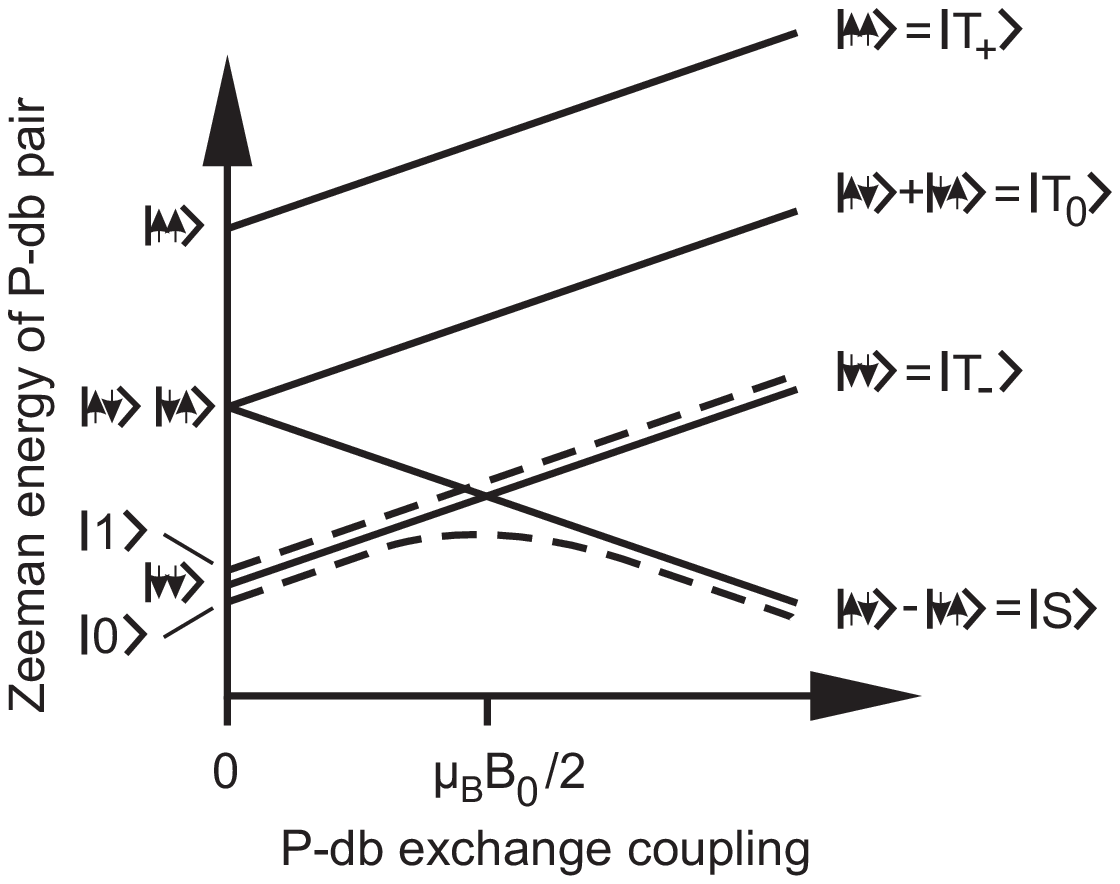}
\end{center}
\caption{\footnotesize{
Adiabatic change of the P-db energy eigenbase. In absence of exchange interaction, the pair 
remains in product base states. When the exchange is turned on, the states change toward 
the singlet or triplet states. At low temperatures, the 
$|T-\rangle$-state is occupied without 
exchange coupling. Due to hyperfine coupling, the change of this state with increasing 
exchange interaction is determined by the state of the nuclear spin of the P donor.}}
\label{fi6}
\end{figure}

The idea behind this setup is to take advantage of Pauli's principle, similar to the 
charge--readout proposed by Kane~\cite{Kane}. 
As long as the R-gate is charged negatively, the wave function overlap between the donor 
electron of the $^{31}$P and the db is small which minimizes a transition probability 
between the two states and keeps the 
spin--exchange negligibly small as well. When the R-gate is charged positively,
such that the wave function overlap between the $^{31}$P and the db is sufficiently 
large, exchange interaction increases. If this increase is introduced 
slowly, an adiabatic change of the spin-pairs' energy eigenstates can take place
from an uncoupled product base into a set of singlet and triplet states.
At low temperatures ($T\leq 100$mK), the uncoupled pair is polarized in a 
$|T-\rangle$-state as long as the coupling is absent. When the exchange is increased slowly, 
this $|T-\rangle$-state remains either unchanged or shifts into a $|S\rangle$-state, 
depending on the orientation of the $^{31}$P-nucleus (illustrated in fig.~\ref{fi6}). 
Note that in presence of a second 
nuclear spin at the db-site, the A-gate above this site would 
have to be charge positively in order to minimize hyperfine coupling. After the R-gate has been  
``opened'' and the electronic pair is in singlet or triplet state, an electronic transition 
can take place which charges the db;  this transition however is possible only
when the electronic pair has singlet content in advance and thus, the charging of the db state 
depends on the nuclear state of the $^{31}$P qbit. 

\begin{figure}[t]
\begin{center}
\includegraphics{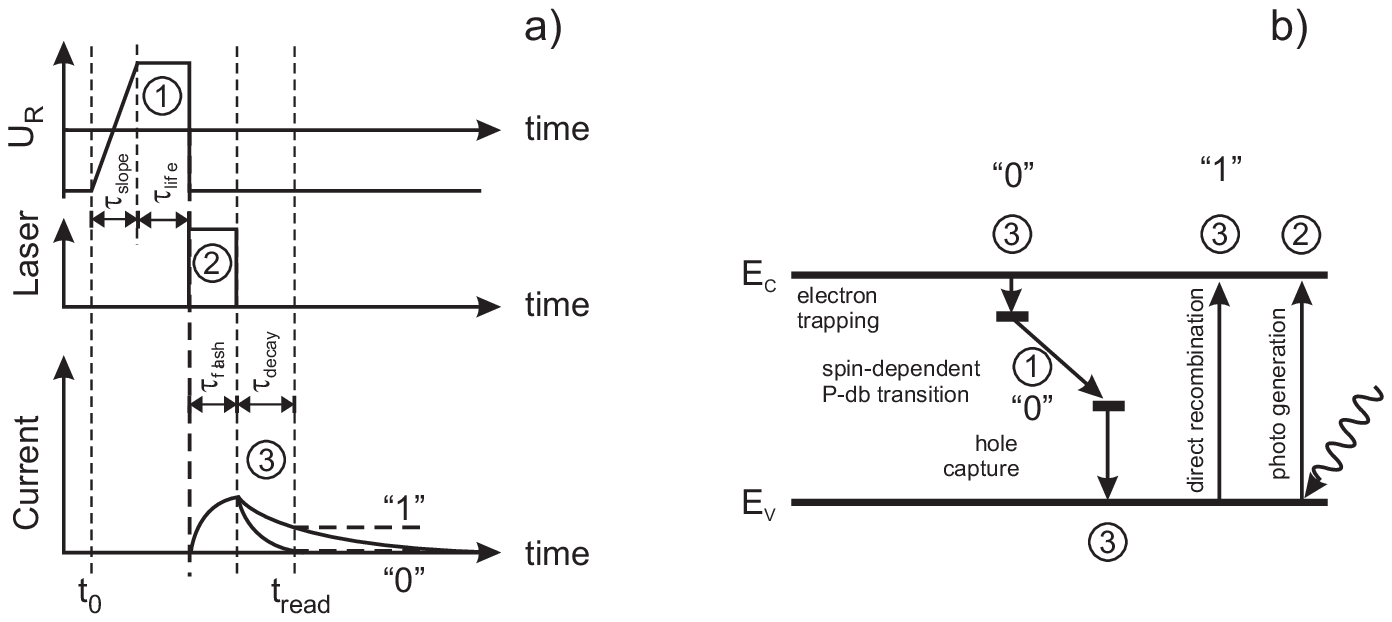}
\end{center}
\caption{\footnotesize{
Sequencial diagram of the readout timing (a) and the corresponding electronic transitions within
the energy levels (b). The different steps of the entire readout process are labelled with 
encircled numbers 1 to 3. At the end of the process at time $t_{\mathrm read}$, a 
photocurrent measurement reveals the value of the qbit. For details see text.
}}
\label{fi7}
\end{figure}

The timing and a scetch of the transitions during a readout sequence are displayed in 
fig.~\ref{fi7}. As long as quantum operations on the qbit take place, no light is imposed 
on the sample and the R-gate is charge negatively. 
When the readout sequence begins at a time $t_0$, the bias of the R-gate is slowly inverted 
towards a positive voltage. Once exchange coupling is established after a time 
$\tau_{\mathrm slope}$, it remains unchanged until the actual recombination transition has 
taken place. The time necessary for this transition is of the order of the electronic 
pair states's lifetime $\tau_{\mathrm life}$, after which the exchange interaction is 
switched off again. The nuclear spin state of the $^{31}$P donor is then coded in the donor 
and db charge state. A short ($\tau_{\mathrm flash}$, ns-Range) and weak (nW-range) 
laser pulse imposed on the sample will then increase conductivity by the generation 
of a few pairs of excess electrons and holes. If the donor and the db state are not 
charged (no transition), a slow decrease of the photoconductivity will follow, which is determined by 
slow band--band recombination in the ultra pure $^{28}$Si bulk. If the two states are 
charged, a fast decay will take place since charge carriers will be 
trapped in the charged states. Thus, the level of the photoconductivity a time 
$\tau_{\mathrm decay}$ after the end of the laser pulse will reveal the result of the readout 
process. Note, that the readout process itself automatically neutralizes the two states such 
that a new series of operations can take place after its completion.

\section{Challenges of the implementation}

The experiments presented above are the motivation for the concept proposed in this study and 
many other questions will have to be answered experimentally before an actual proof of this concept 
can be given. The implementation of the singly occupied deep level state will mostly 
depend on whether it is possible to control its location and whether its charge 
carrier capture cross sections and the given transition times of the system are fast enough.
The dangling bond present in amorphous and microcrystalline silicon would be an ideal system 
with regard to the latter properties. However, the high disorder in the 
microcrystalline morphology of silicon makes it a bad choice for QC. Since no process has 
been established which allows the creation of a single db with $\AA$-accuracy at an arbitrary 
site, a different way of deep level implementation must be chosen.

Various impurities and defects provide singly occupied, paramagnetic deep level donor states 
in c-Si, some of which are gold, potassium, strontium or chromium~\cite{sze}. 
An additional possibility could be interface recombination. 
Recombination at c-Si/SiO$_2$--interfaces has shown to have 
spin-dependent paths~\cite{stathis}. Hence, the very same processes which limited the original pure
c-Si QC-concept and led to the proposal of Si/SiGe-heterostructures~\cite{Kane2} 
could actually be benificial for a recombination readout mechanism. 

Beside the questions for a proper implementation of the deep level center, several other issues 
remain to be investigated: 
Due to the necessity of error correction, one of the preconditions for QC is that 
readout does not destroy the state of the qbit. Therefore, only P-deep level--transitions 
are possible which leave the $^{31}$P--relaxation unchanged. Important for the feasibility 
of the recombination readout will also be the question of the overall readout time which 
should not exceed a lower microsecond range. When the P-deep level transition takes place, 
the slow relaxation of non--equilibrium polarization at low temperatures could pose a problem 
due to charged P-deep level--pairs which are neutralized by excess charge carriers produced in 
excited spin states. The time needed  for the relaxation of these states could be very long. 
However, this problem may be solved by injection of excess carriers by spin--polarized 
electronic injection instead of light injection. 
Finally, the question for the one--qbit sensitivity has to be asked: 
Time--domain measurements of spin--dependent recombination
have shown to be sensitive enough to detect the influence of as little as 50 charge carrier pairs, 
even though the experiment was carried out in the presence of a strong constant photocurrent offset.
For the readout as proposed, a measurement without offset could be made and the detection would be
even more sensitive. A 2eV laser pulse of 1ns length and 3.2nW intensity produces 10 electron--hole 
pairs if the internal quantum efficiency is assumed to be 100\%. The detection of these 10 charge 
carrier pairs would require a pA--current measurement on a microsecond scale which does not pose a 
problem -- however, whether the 10\% difference between the two readout results caused by the one 
recombination center are detectable can only be proven when an implementation of the proposed 
setup is available.

\section{Summary}

The demonstration that coherent spin--states of electrons and holes trapped in CE and db states can 
govern their recombination rate has motivated the idea of a readout mechanism for $^{31}$P qbits in 
c-Si. A concept, based on a sequence of exchange gating, light pulse and photocurrent 
measurements was proposed, various implementations suggested and possible limitations and drawbacks 
discussed.
\clearpage
\bibliographystyle{prsty}

\end{document}